\documentclass[%
 reprint,
 amsmath,amssymb,
 aps,
prl,
]{revtex4-2}

\usepackage{graphicx}% Include figure files
\usepackage{dcolumn}% Align table columns on decimal point
\usepackage{bm}% bold math
\usepackage{color,soul,xcolor}

\begin{document}

\preprint{APS/123-QED}

\title{Mechanism of stochastic resonance in viscoelastic channel flow}% Force line breaks with \\
%\thanks{A footnote to the article title}%

\author{Yuke Li}
 \email{yuke.li@weizmann.ac.il}
 %\altaffiliation{Department of Physics of Complex Systems, Weizmann Institute of Science, 234 Herzl Street, POB 26, Rehovot 7610001 Israel.}%Lines break automatically or can be forced with \\
\author{Victor Steinberg}
 \email{victor.steinberg@weizmann.ac.il}
\affiliation{%
 Department of Physics of Complex Systems, Weizmann Institute of Science,  Rehovot 7610001, Israel
}%

%\collaboration{MUSO Collaboration}%\noaffiliation
\date{\today}% It is always \today, today,
             %  but any date may be explicitly specified

\begin{abstract}

We have recently discovered stochastic resonance (SR) in chaotic inertia-less viscoelastic channel flow. SR appears just above a pure elastic instability at a critical Weissenberg number, $Wi_c=150$, of a transition regime. In this lower sub-region up to $Wi\sim 300$, only the streamwise velocity, $u$, exhibits a chaotic spectrum, $E_u$, while the spanwise velocity continues to exhibit white noise, verified by its flat spectrum, $E_w$, accompanied by weak intensity elastic waves. However, SR vanishes at the upper limit at $Wi\sim 300$, when $E_w$ becomes chaotic, indicating $Wi$ as the control parameter. Here we clarify the mechanism of SR emergence by validating the control parameters, namely $Wi$ and the rms velocity fluctuations, $u_{rms}$, measured at multiple channel locations, which determine the range of SR existence. Our experiments verify three key ingredients of the SR mechanism: chaotic $E_u$, white noise $E_w$, and weak elastic waves, which are consistent with three constituents of autonomous dynamical systems exhibiting SR.

\end{abstract}
%\keywords{viscoelastic channel flow, elastically driven instability, supercritical non-modal instability, stochastic instability mechanism}
\maketitle

%%\label{sec:intro}

%Recently, we disclosed the route from laminar to sustained chaotic flow in inertia-less viscoelastic channel flow externally perturbed by finite-amplitude white noise \cite{yuke2023preprint}. In the limited range above the onset of supercritical, non-normal-mode elastic instability, $Wi_c$, in a lower subrange of a transition flow regime, the laminar flow becomes chaotic only in the streamwise direction, $x$, while in the spanwise direction, $z$, the flow continues to exhibit flat white noise spectrum of the spanwise velocity, $w$, and very low intensity peaks of elastic waves. Moreover, in the lower subrange, the time series of the streamwise velocity, $u(t)$, shows nearly periodic spikes on the top of the chaotic flow. These spikes are characterized by a narrow, low frequency, high energy peak on the top of the chaotic streamwise velocity power spectrum, $E_u$. Here, $Wi= U\lambda/ d$ is the main control parameter in inertia-less viscoelastic fluid flows that denotes the ratio of elastic stress to relaxation dissipation of elastic stress, where $U$ is the average flow velocity defined by fluid discharge rate (see the Supplemental Material \cite{SM} and also references \cite{solutionpreparation,OpenPIV} therein), $d$ is the characteristic vessel size, and $\lambda$ is the longest polymer relaxation time. For inertia-less, viscoelastic channel flow above $Wi_c$, Reynolds number $Re=Ud \rho/\eta \ll 1$ and $Wi\gg1$, where $\rho$ and $\eta$ are the solution density and dynamic viscosity, respectively. 

In several recent publications we have reported a direct transition from laminar to sustained chaotic flow at $Wi>Wi_c$ and $Re\ll 1$ in inertia-less viscoelastic channel flow. It occurs via a supercritical non-normal mode elastic instability at $Wi_c$, due to finite-amplitude external white noise perturbations $u_{rms}$ at the inlet $l/h=0$, despite the proven linear stability of viscoelastic parallel shear flows \cite{Leonov_JMM1967,Renardy_JNFM1986}. Here, the Weissenberg number $Wi= U\lambda/ d$ is the main control parameter in inertia-free viscoelastic fluid flow, denoting the ratio of the elastic stress to its relaxation dissipation, where $U$ is the mean flow velocity defined by the fluid discharge rate (see the Supplementary Material \cite{SM} and the references \cite{solutionpreparation,OpenPIV} therein), $Wi_c=150\pm 10$ is the instability onset for the discussed channel, $d$ is the characteristic vessel size, and $\lambda$ is the longest polymer relaxation time, $u_{rms}$ is a root mean square (rms) of the streamwise velocity fluctuations, $l$ is the location along the channel, and $h$ is the channel height. 

%For inertial-free viscoelastic channel flow above $Wi_c=150\pm 10$, the Reynolds number $Re=Ud \rho/\eta \ll 1$, where $\rho$ and $\eta$ are the solution density and dynamic viscosity, respectively.

% Moreover, in our previous work, a viscoelastic channel flow with strong pre-arranged perturbations at $l/h=0$, excited by an obstacle array that generates a chaotic flow different from white noise, also shows universal scaling  with $Wi-Wi_c$ in the transition regime \cite{jha2020universal}. Thus, we conclude that for a wide range of $Wi>Wi_c$ in three flow regimes, namely transition, elastic turbulence (ET), and drag reduction (DR) accompanied by elastic waves, the flow behavior is similar regardless of the $u_{rms}$ value at the inlet \cite{Li2023non-Hermitian,Li2023flowprop}. 

The next significant step in uncovering a mechanism to promote chaotic channel flow at $Wi>Wi_c$ in three flow regimes is the proposed and experimentally verified mechanism of vortex amplification by elastic waves \cite{Li2023flowprop}. Therefore, the key role of elastic waves in amplifying the wall-normal vortex fluctuations is the resonant pumping of elastic wave energy extracted from the mean shear flow into fluctuating vortices at $Wi>Wi_c$, which requires sufficient elastic wave energy and could be problematic near $Wi_c$ where elastic waves are just initiated. 
Thus, the goal of our recent study \cite{yuke2024preprint} was to uncover the transition to chaotic flow near $Wi_c$. We increase the resolution near threshold to study flow behaviours near $Wi_c$. First, this study reveals two sub-regions at $Wi>Wi_c$ in the transition regime: lower at $150<Wi\leq 300$ and upper at $400\leq Wi\leq 1000$. In the lower sub-region, the laminar flow becomes chaotic only in the streamwise direction, $x$, while in the spanwise direction, $z$, the spanwise velocity, $w(t)$, power spectrum, $E_w$ remains flat indicated white noise fluctuations as at $Wi<Wi_c$ and with low intensity peaks of the elastic waves. In addition, the time series of the streamwise velocity $u(t)$ and pressure $p(t)$ show large and nearly periodic negative spikes. These spikes result in low-frequency, high-energy peaks in the streamwise velocity, $E_u$, and pressure, $E_p$, power spectra.  As suggested in Ref. \cite{yuke2024preprint}, the nearly periodic sharp low-frequency and high-energy  peaks in $E_u$ and $E_p$ in the lower sub-region resemble stochastic resonance (SR).  Moreover, high-energy elastic waves now take over the key role from SR, since they are able to amplify the wall-normal vorticity fluctuations and self-organize the streaks into cycles synchronized at the elastic wave frequency, $f_{el}$ \cite{Li2023flowprop}. Thus, SR is the key factor of the stochastic path to the persistent chaotic flow just above $Wi_c$ in the lower sub-region. However, this finding is based only on the study of viscoelastic channel flow at the fixed spatial location $l/h=380$ \cite{yuke2024preprint}. To our knowledge, this marks the first experimental observation of SR in chaotic systems within fluid dynamics.

 %while in the upper sub-region the periodic negative spikes disappear along with low-frequency, high-energy peaks in both $E_u$ and $E_p$, while both $E_u$ and $E_w$ become chaotic. As suggested in Ref. \cite{yuke2024preprint}, the nearly periodic sharp low-frequency and high-energy  peaks in $E_u$ and $E_p$ in the lower sub-region resemble stochastic resonance (SR).
 
In the upper sub-region at $400\leq Wi\leq 1000$, the SR disappears along with the appearance of both $E_u$ and $E_w$ as chaotic power spectra and high intensity elastic waves. Thus, the low-frequency, high-energy SR detected in the lower sub-region ensures the transition from laminar to sustained chaotic flow above $Wi_c$. This is reminiscent of a scenario that  occurs in an autonomous dynamical chaotic system interacting with external white noise in the presence of weak periodic modulation \cite{anishch1993jstatphys,anishch1999phys_uspekhi}. It differs fundamentally from classical SR, which involves either weak oscillations with external white noise or a bistable system driven by both external noise and a weak periodic signal \cite{benzi1981JPhys,Hanggi1998revmodphys}. In deterministic chaotic systems, quasi-attractors merge into a single chaotic attractor above a critical parameter value in an event known as an attractor crisis \cite{anishch1999phys_uspekhi}. Post-crisis, the merged attractor intermittently switches in a ``chaos-chaos" manner, forming a ``deterministic stochastic attractor" interacting with external noise. This switching can be synchronized by weak periodic modulation, producing SR at a fixed control parameter value \cite{anishch1993jstatphys,anishch1999phys_uspekhi}. 

Here we present experimental results obtained in the same inertia-less viscoelastic channel flow and with the same approach as in Ref. \cite{yuke2024preprint}, where SR is discovered and characterized in the lower sub-region of the transition regime, but at the multiple spatial locations $l/h$ along the channel length, where $l/h$ plays the role of the second controlled parameter besides $Wi$. 

%The main goal of the experiment is to find out whether the upper bound of the lower sub-region for $Wi$ remains constant along the channel or depends on $l/h$. If such a dependence exists, we will reveal a role of the second control parameter, $l/h$, in defining the range of SR existence along the channel at a given value of $Wi$.

The main goal of the experiment is to verify the mechanism of SR emergence and to validate the control parameter besides $Wi$, defining the range of SR existence. Moreover, we intend to reveal the relation between elastic wave  and SR frequencies, important feature to verify the proposed mechanism of SR appearance.
 
 %namely $Wi$ and the rms velocity fluctuations, $u_{rms}$, measured at multiple channel locations. Both $Wi$ and $u_{rms}$ determine the range of SR existence.

%We conducted the experiment in a long channel of dimensions 500 ($L$, $x$: streamwise) $\times$ 0.5 ($h$, $y$: wall-normal) $\times$ 3.5 ($W$, $z$: spanwise) mm$^3$ (see Fig.S1 in \cite{SM}), the same as used in Ref. \cite{yuke2023preprint}. The channel has two sources to initiate finite-size perturbations, namely a non-smoothed inlet and two holes of 0.5 mm diameter in the top plate close to the inlet and outlet, used for pressure measurements. A dilute aqueous solution of 64\% sucrose concentration and 80 ppm of high molecular weight polyacrylamide (PAAm) is the same as used in Ref. \cite{yuke2023preprint}. The longest polymer relaxation time ($\lambda$) is 13 s. The average velocity $U$, used in both $Wi$ and $Re$, is obtained via the flow discharge rate, $Q$, as $U = Q/\rho Wh$.  In SM Methods, we present the details of the experimental setup, preparation and characterization of polymer solution, and Particle Image Velocimetry (PIV) technique to measure 2D velocity field in $x-z$ horizontal plane. 

We performed the experiment in a long channel of dimensions 500 ($L$, $x$: streamwise) $\times$ 0.5 ($h$, $y$: wall-normal) $\times$ 3.5 ($W$, $z$: spanwise) mm$^3$ in Fig. S1 in \cite{SM}. The channel has two sources of finite size perturbations, namely an unsmoothed inlet and two 0.5 mm diameter holes in the top plate near the inlet and outlet used for pressure measurements. A dilute aqueous solution of 64\% sucrose concentration and 80 ppm high molecular weight polyacrylamide (PAAm) is the same as that used in Ref. \cite{yuke2024preprint}. The longest polymer relaxation time $\lambda$ is 13 s. The average velocity $U$, used in both $Wi$ and $Re$, is obtained via the flow discharge rate, $Q$, as $U = Q/\rho Wh$.  In the Methods in the Supplementary Material \cite{SM}, we present the details of the experimental setup, the preparation and characterization of the polymer solution, and the Particle Image Velocimetry (PIV) technique to measure the 2D velocity field in the $x-z$ horizontal plane.

\begin{figure}
  \includegraphics{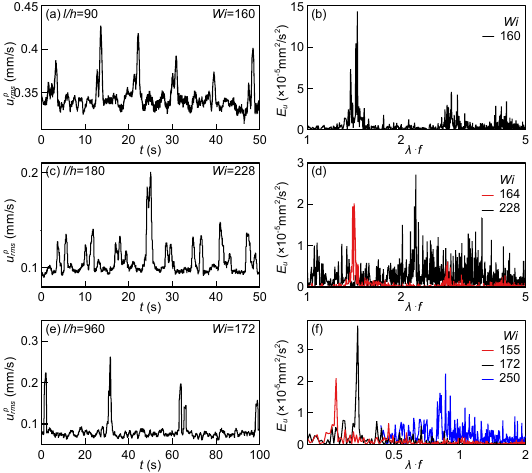}% Images in 100% size
  \caption{\label{fig1:Eu} Partially averaged over each consecutive 1s (100 samples) stream-wise velocity fluctuations, $u^{p}_{rms}(t)$, and $u$ energy spectra, $E_{u}$, at various $l/h$ and $Wi$ values. 
  At $l/h$=90: (a) $u^{p}_{rms}(t)$ at $Wi$=160 and (b) $E_{u}$ at $Wi$= 160;
  at $l/h$=180: (c) $u^{p}_{rms}(t)$ at $Wi$=228 and (d) $E_{u}$ at $Wi$= 164 and 228; 
  at $l/h$=960: (e) $u^{p}_{rms}(t)$ at $Wi$=172 and (f) $E_{u}$ at $Wi$= 155, 172 and 250.
  }
\end{figure}

In Fig. \ref{fig1:Eu}, the left column shows three plots of the temporal evolution of $u^p_{rms}(t)$, the partially averaged over consecutive 1s (100 samples) streamwise velocity fluctuations, at three positions, $l/h$: (a) 90 at $Wi=160$; (c) 180 at $Wi=228$; (e) 960 at $Wi=172$. The right column shows $u(t)$ power spectra, $E_u$, in lin-log presentation at the corresponding $l/h$ and various $Wi$ values in the lower sub-region of the transition regime: (b) 90 at $Wi=160$; (d) 180 at $Wi=164,228$; (f) 960 at $Wi=155,172,250$. In Figs. \ref{fig1:Eu}(a),(c),(e), the $u^p_{rms}(t)$ time evolution shows almost periodic spikes of the partially smoothed data, whose $E_u$ in Figs. \ref{fig1:Eu}(b),(d),(f) show sharp, high energy peaks associated with SR. 

\begin{figure}
  \includegraphics{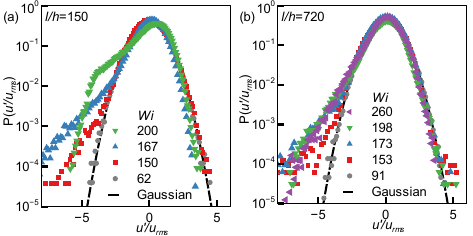}% Images in 100% size
  \caption{\label{fig2:PDF} Probability density functions $P(u'/u_{rms})$, at different $l/h$ and $Wi$ s: 
  (a) at $l/h$=150 and $Wi$=62, 150, 167 and 200; and 
  (b) at $l/h$=720 and $Wi$=91, 153, 173, 198 and 260. 
  The dashed black lines represent the Gaussian PDF.}
\end{figure}

\begin{figure}
  \includegraphics{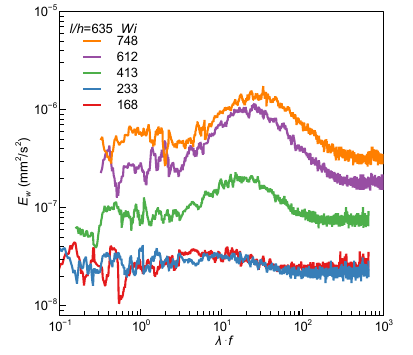}% Images in 100% size
  \caption{\label{fig3:Ew} Spanwise velocity energy spectra in log-log scales measured at the channel center far from the inlet: $E_{w}$ at $l/h$=635 and $Wi$=168, 233, 413, 612 and 748.}
\end{figure}

\begin{figure}
  \includegraphics{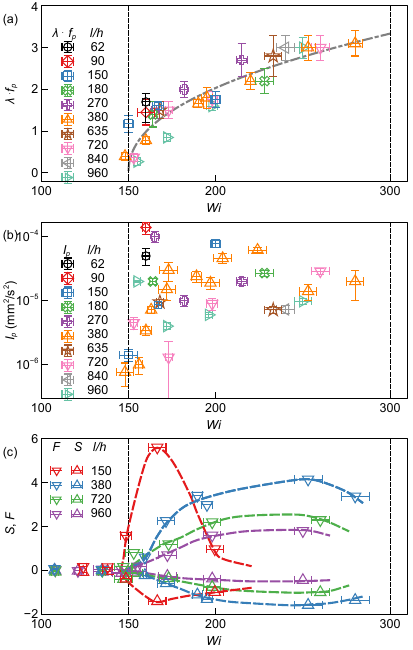}
  \caption{\label{fig4:SF}Properties of spikes near threshold instability in two sub- regions. 
  (a) Normalized SR frequency, $\lambda\cdot{f}_{p}$, versus $Wi$ at different $l/h$. The black dashed line is the power-law fit $\lambda\cdot{f}_{p}\sim (Wi-Wi_c)^{\gamma}$ with $\gamma=0.45\pm 0.05$. (b) SR peak energy $I_{p}$ in $E_u$ versus $Wi$ at the same $l/h$. (c) Skewness, $S$, and flatness (kurtosis), $F$, of $P(u'/u_{rms})$ versus $Wi$ at four $l/h$ values. The colored dashed lines show trends. The vertical black dashed lines in (a-c) mark $Wi_c=150$ and $Wi$=300, as  limits of the lower sub-region.
  }
\end{figure}

In Fig. \ref{fig2:PDF} we show PDFs of $P(u'/u_{rms})$ at $l/h=150, 720$ and several $Wi$ values in the lower sub-region of the transition regime. As in ref.\cite{yuke2024preprint}, strong deviations from the Gaussian PDF appear for negative values of $u\prime/u_{rms}$ for all $Wi$. It is noteworthy that further down, the deviations of the left tail from the Gaussian PDF become closer to exponential than another shifted Gaussian PDF, as observed at $l/h=380$ in Fig. 2 in Ref. \cite{yuke2024preprint}. The characterization of the PDF shape deviation from Gaussian is quantified by its third and fourth moments, called skewness, $S$, and  flatness (kurtosis), $F$, respectively, shown in Fig. \ref{fig4:SF}(c) as a function of $Wi$ values in the lower sub-region over a wide range of $l/h$. At $l/h=380,720,960$, sufficiently far from the inlet, both $S$ and $F$ exhibit a non-monotonic dependence on  $Wi$ with a smoothed maximum in the middle of the lower sub-region, so that the absolute values of $S$ and $F$ decrease with increasing $l/h$. Meanwhile, at $l/h=150$, $F$ grows abruptly above $Wi_c$ and then drops significantly towards $Wi\approx 200$.

As mentioned in the introduction and also in Ref. \cite{yuke2024preprint}, SR is observed only in the lower sub-region at $l/h=380$, where the remaining flat spectrum $E_w$, which denotes white noise versus $\lambda f$ in log-log coordinates, shows a weak growth with $Wi$. According to the proposed similarity of the SR appearance in the autonomous dynamical chaotic system \cite{anishch1993jstatphys,anishch1999phys_uspekhi}, the interaction of the chaotic attractor with external white noise in the presence of weak oscillations is the necessary condition for the SR appearance. To verify the dependence of $E_w$ on $\lambda f$ at several positions, $l/h$ and various $Wi$ in the lower sub-region of the transition regime are plotted in log-log coordinates for $l/h=635$ and $Wi=168,233,413,612,748$ in Fig. \ref{fig3:Ew} and additionally for $l/h=22,62,90,180,840$ and various $Wi$ values in Fig. S3 \cite{SM}. The results shown in Fig. \ref{fig3:Ew} and Figs. S3(c),(d),(e) of \cite{SM} at $l/h$ from 90 to 840 confirm the observation reported in Ref. \cite{yuke2024preprint} that SR is observed only for chaotic $E_u$ interacting with a flat white noise $E_w$ accompanied by weak elastic waves.  However, near the inlet at $l/h=22$, $E_w$ remarkably exhibits chaotic spectra for all $Wi$ values throughout the transition regime in Fig. S3(a) \cite{SM}. Similarly, at $l/h=62$ and 90 $E_w$ remains chaotic except at $Wi=162$, where $E_w$ is nearly flat. Thus, the SR exists only in the lower sub-region where the flat, white noise $E_w$ is found, as shown by the phase diagram in $Wi$ and $l/h$ coordinates (Fig. S4 \cite{SM}). This is the key result first reported at $l/h=380$ in Ref. \cite{yuke2024preprint}. 

%To compare $I_p$ with $I_{el}$, in the spanwise velocity power spectra, $E_w$, observed at $Wi>Wi_c$, we present $E_w$ versus $\lambda f$ in log-log coordinates at $l/h=635$ and various $Wi$ values, covered the whole transition regime, in Fig. \ref{fig3:Ew} and for $l/h=90,180,840$ and various $Wi$ values in SM Fig. S3(a),(b),(c). In the lower subrange at $l/h=380$, one of the findings reported in \cite{yuke2023preprint} is a slow growth of $E_w$, which remains flat denoting white noise spectrum, except for at $l/h=22,62,90$, where $E_w$ remarkably indicates chaotic spectrum. Further on, $I_{el}$ stays up to $~4\times 10^{-7}(mm^2/s^2)$, close to the resolution limit. Here in SM Table 1 and Fig. S5(b), we present similar results on $I_{el}$, and the normalized elastic wave frequency, $\lambda f_{el}$, in SM Fig. S5(a) obtained in both subranges of the transition regime at various $Wi$ values and ten locations, $l/h=$ 62, 90, 150, 180, 270, 380, 635, 720, 840, 960. In the lower subrange, $I_{el}$ remains flat and slow increases up to the upper bound at $8\times 10^{-7}(mm^2/s^2)$, marked by red dotted line for all positions, in SM Fig. S5(b). 
%

One reason for the dependence of the upper $Wi$ bound of the lower sub-region on $l/h$, where SR is detected, is the fact that $u_{rms}/U$ and the normalized elastic wave energy, $I_{el}/w^2_{rms}$, decrease with $l/h$ at all $Wi$ in the transition regime. It was first observed in a channel with six small holes in a top plate, shown in Fig. 6 of Ref. \cite{Li2023non-Hermitian} and Fig. S2 in \cite{SM} for the current channel. First, since both $u_{rms}/U$ and $I_{el}/w^2_{rms}$ persist up to the channel outlet, the non-normal elastic instability due to continuous perturbations at the inlet is absolute (or global), as first pointed out in Ref. \cite{Li2023non-Hermitian}. This means that convective (or local) instability is absent, in contrast to Newtonian open flows where an initial convective instability followed by absolute instability is theoretically explained in  refs. \cite{barkley2016,huerre1990ARFM,chomaz2005ARFM} and experimentally confirmed in refs. \cite{tsameret1991EPL,tsameret1991PRL,babcock1991PRL,babcock1992PhysD,tsameret1994PhysRevE}. As already suggested in Ref. \cite{Li2023non-Hermitian}, these discrepancies between the properties of Newtonian and viscoelastic open flows are caused by the different nature of the nonlinear terms: Only one advection term in the velocity field in the former versus three nonlinear terms, including one advection term, in the elastic stress field in the latter. 
Moreover, in contrast to Newtonian open flows, the viscoelastic flow properties depend on $l/h$ above $Wi_c$ in Figs. S2(a), (b) in \cite{SM}. Furthermore, the $u_{rms}/U$ dependence on $l/h$ is the same for the lower and upper sub-regions in Fig. S2(a) in \cite{SM}, while the $I_{el}/w^2_{rms}$ dependence on $l/h$ is different in both sub-regions in Fig. S2(b) in \cite{SM}. Furthermore, both $u_{rms}/U$ and $I_{el}/w^2_{rms}$ show two distinct regions with similar $l/h$ dependencies: a steep decrease near the inlet and a gentle decrease further away. Near the inlet, the steep decrease of  $u_{rms}/U$ due to external perturbations occurs from about $14\%$ down to $\sim 4\%$ at $l/h\approx 90$ and then decreases to $\sim\ 3\%$ near the outlet. While for $I_{el}/w^2_{rms}$ the transition locations from steep to gentle slope occur for the lower sub-region at $\sim 140$ and for the upper sub-region at $l/h\approx 90$, and as a result the steepness for $I_{el}/w^2_{rms}$ differs for the lower and upper sub-regions far from the inlet in Fig. S2(b) in \cite{SM}. To summarize, the observed dependence of the SR existence range and the elastic wave intensity on $l/h$ is due to the dependence of $u_{rms}/U$ on $l/h$. This means that $u_{rms}/U$ is the second control parameter besides $Wi$. Thus, one could expect that different values of $u_{rms}/U$ at the inlet would also affect the presence range of SR and flat white noise $E_w$.

%As already pointed out in Ref. \cite{Li2023non-Hermitian}, these discrepancies between the properties of Newtonian and viscoelastic open flows are caused by the different nature of the nonlinear terms: Only one advection term in the velocity field in the former versus three nonlinear terms, including one advection term, in the elastic stress field in the latter.

In addition, the SR energy peak, $I_{p}$, and the SR normalized frequency, $\lambda f_p$, observed exclusively in the lower sub-region of the transition regime at various $Wi$ values and ten locations, $l/h=$ 62, 90, 150, 180, 270, 380, 635, 720, 840, 960, are shown in Figs. \ref{fig4:SF}(a),(b) and also provided in {Table S1 \cite{SM}}. The SR normalized frequency, $\lambda f_p$, grows with $Wi$ at all locations, and the fit of all data shows a scaling relationship $\lambda f_p\sim(Wi-Wi_c)^{\gamma}$ with $\gamma=0.45\pm0.05$, the same as at $l/h=380$ in Ref. \cite{yuke2024preprint}. The $I_p$ shows mostly growth with $Wi$ for all $l/h$ values, and then a slight decrease towards $Wi\approx 300$, the upper limit of the lower sub-region.

In Figs. S5(a), (b) \cite{SM} we plot $\lambda f_{el}$ and $I_{el}$ as functions of $Wi$, at eleven locations $l/h=$22, 62, 90, 150, 180, 270, 380, 635, 720, 840, 960 in both sub-regions of the transition flow regime. In the lower sub-region, $E_w$ remains flat as discussed above, and $I_{el}$ slowly increases up to the upper limit at $8\times 10^{-8}(mm^2/s^2)$, marked by the red dotted line for all positions where $I_{el}$ has two orders of magnitude lower energy than $I_p$ ( Fig. \ref{fig4:SF} and Fig. S5(b) \cite{SM}). Only for $l/h=22$ and $l/h=62$ do the $I_{el}$ values find an order of magnitude larger than the upper limit at $Wi=162,235$ and $Wi=162$, respectively, and remain larger than those at the other locations, although they vary at about the same rate as for the other $l/h$ values. In Fig. S5(a) \cite{SM} one finds the growth of $\lambda f_{el}$ with $Wi$ at all positions along the channel, although at $l/h=22,62,90$ $\lambda f_{el}$ remains smaller than the rest in both sub-regions. To compare the scaling of $\lambda f_p$ and $\lambda f_{el}$ versus $Wi-Wi_c$, we plot $\lambda\cdot{f_{el}}\sim(Wi-150)^{\beta}$ with $\beta=0.4\pm 0. 1$ in the lower sub-region at eleven locations $l/h=$22, 62, 90, 150, 180, 270, 380, 635, 720, 840, 960 (Fig. S5(a) \cite{SM}), which is close to the exponent $\gamma=0.45\pm 0.5$ obtained for $\lambda f_p$ in Fig. \ref{fig4:SF}. In the upper sub-region, $E_w$ becomes chaotic with the power-law decay at high frequencies (see Fig. \ref{fig3:Ew} and Figs. S3(a),(b),(c)\cite{SM}), and SR disappears. The latter is accompanied by a steep growth of $I_{el}$, reaching the $I_p$ values observed in the lower sub-region.
 Moreover, $E_w$ remains flat and grows up to $\sim 10^{-7}(mm^2/s^2)$, close to the resolution limit {Table S1 \cite{SM}}.   
 
%Summarizing, we observe a strong dependence of the $Wi$ upper bound of the lower subrange of the SR appearance as a function of $l/h$ close to the inlet, where $u_{rms}/U$ is up to 4 times larger at $l/h=22$ for $Wi=162$ and 235, and at $l/h=62$ for $Wi=237$ about twice larger than far away from the inlet. Such large $u_{rms}/U$ at these values of $l/h$ causes chaotic $E_w$ and much larger $I_{el}$ values even for $Wi$ slightly larger $Wi_c$. For such reasons, both the SR and lower subrange are absent at $l/h=22$, and for $l/h=62$ the upper bound is located up to $W\leq237$, lower than $Wi\geq 300$, the upper bound for larger $l/h$ values. It means that the existence range of SR depends on $u_{rms}$ value, which determines whether $E_w$ is flat, noting white noise spectrum, or chaotic, when the mechanism of SR excitation is ceased to be effective. To verify the connection between the elastic waves and SR, we found that the exponents $\beta$ and $\gamma$ in the scaling relations of $f_{el}$ and $f_p$ on $Wi-Wi_c$ at all $l/h$ have close values $0.45\pm0.5$ and $0.4\pm0.1$, respectively, inside the error bars that evidences their direct relation. However, the range  of $\lambda f_p$ is significantly shifted down for four-five times compared to the existence range of $f_{el}$ (Fig. \ref{fig4:SF}(a) versus {Fig. S5(a) \cite{SM}), that is contrary to the dynamical model with $f_p$ shifted up relatively to $f_{el}$, and its value is determined by the noise strength \cite{haken1993prl,Ning1994pre}. 

In summary, we observe a strong dependence of the $Wi$ upper bound of the lower sub-region as a function of $l/h$ near the inlet. At $l/h=22$ for $Wi=162$ and 235, $u_{rms}/U$ is up to 4 times larger, and at $l/h=62$ and $Wi=235$, $u_{rms}/U$ is about twice as large as far downstream. Such large $u_{rms}/U$ values near the inlet lead to chaotic $E_w$ and much larger $I_{el}$ values either at $l/h=22$ for all $Wi>Wi_c$  or at $l/h=62$ and 90 in the upper limit at $Wi=162$, close to $Wi_c$. This means that the range of existence of SR and flat white noise $E_w$ depends on the value of $u_{rms}/U$. To verify the relation between SR and elastic waves, we show that the exponents $\beta$ and $\gamma$ in the scaling relations of $f_{el}$ and $f_p$ on $Wi-Wi_c$ have close values of $0.45\pm 0.5$ and $0.4\pm 0.1$ within the error bar, proving their direct relation. However, the range  of $\lambda f_p$ is significantly shifted down four to five times compared to the range of $f_{el}$ (Fig. \ref{fig4:SF}(a) versus (Fig. S5(a) \cite{SM}), which is in contrast to the dynamical chaotic model with $f_p$ shifted up relative to $f_{el}$, and its value is determined by the noise strength \cite{anishch1993jstatphys,anishch1999phys_uspekhi}.

In conclusion, SR appears exclusively in the lower sub-region of the transition flow regime just above $Wi_c$. There are three key factors  that define the range of SR existence: (i) the presence of flat white noise $E_w$, (ii) low-intensity elastic waves, and (iii) the presence of $u_{rms}/U$, the second control parameter besides $Wi$. When $E_w$ becomes chaotic, SR ceases to exist because the mechanism of SR generation is no longer effective. The next crucial observation is the same scaling dependence of both frequencies, $f_p$ and $f_{el}$, on $Wi-Wi_c$, indicating the direct relationship between SR and elastic waves, although $f_p$ is shifted down 4-5 times relative to $f_{el}$. Thus, these two basic observations are the critical ingredients that allow the use of the autonomous dynamical model, where the chaotic attractor interacts with external noise in the presence of limit cycles (oscillations) leading to SR, with the experimental findings in the much more involved viscoelastic channel flow. We have to note that our experimental observations are obtained in a flow with infinite  degrees of freedom, in contrast to the autonomous dynamical chaotic model \cite{anishch1993jstatphys,anishch1999phys_uspekhi}.

In the lower sub-region above $Wi_c$, SR takes the role of elastic waves whose energy peak $I_{el}$ is more than two orders lower than $I_p$. As a result, SR initiates the appearance of random streaks and wall-normal vortex fluctuations by transferring energy from the main flow to the vortex fluctuations. When $E_w$ becomes chaotic at higher $Wi$, $I_{el}$ also increases significantly to take over the role of a driving factor to promote self-organization of streak cycles synchronized by the elastic wave frequency, $f_{el}$, and to amplify wall-normal vortex fluctuations via the mechanism previously described and verified by us \cite{Li2023flowprop}.

%The described mechanism of SR excitation reminds the feedback response of the viscoelastic channel flow on the emergence above the elastic instability onset of the spanwise velocity flat power spectrum characterized white noise and the low intensity elastic waves, which are  

%\backsection[Supplementary data]{\label{SupMat}Supplementary material and movies are available at \\https://doi.org/10.1017/jfm.2019...}
%\begin{acknowledgments}
We thank Guy Han and Rostyslav Baron for their help with the experimental setup. This work was partially supported by the Israel Science Foundation (ISF, grant \#784/19). 
%\end{acknowledgments}

%Q\nocite{*}
\bibliography{bibliography}

\end{document}

% For the title section we want to reproduce the title section of the Problem Set and add your names.

\thispagestyle{empty} % This command disables the header on the first page. 

\begin{tabular}{p{15.5cm}} % This is a simple tabular environment to align your text nicely 
{\large \bf Supplementary Materials} \\
for the article \enquote{ Mechanism of stochastic resonance in viscoelastic channel flow}\\ Yuke Li, Victor Steinberg*  \\ Department of Physics of Complex Systems, Weizmann Institute of Science, Rehovot 7610001, Israel 
\\victor.steinberg@weizmann.ac.il
*\\
\hline % \hline produces horizontal lines.

\end{tabular} % Our tabular environment ends here.

\vspace*{0.3cm} % Now we want to add some vertical space in between the line and our title.

\subsection*{Experimental setup and Methods}

The experiments are carried out in a straight channel with a rectangular cross-section, made of acrylic and screwed to a metal frame. The dimensions of the channel are 500 mm in the streamwise direction (\textit{L}, x), 3.5 mm in the spanwise direction (W, z) and 0.5 mm in the vertical (wall-normal) direction (\textit{h}, y), giving a total volume of $500\times 3.5\times 0.5$ $mm^3$, as shown in Fig. S1. The polymer solution for the experiment is stored in a large stainless steel container and is driven by compressed nitrogen gas, controlled by pressure regulators. As the fluid exits the container, it first passes through two right-angled bends before flowing into the channel and is then collected in a glass container mounted on a balance (BPS-1000-C2-V2, MRC) at the outlet. The flow discharge is measured instantaneously as a function of time, $m(t)$, using the PC-interfaced balance. The time-averaged fluid discharge rate is calculated as $Q = \langle\Delta m/\Delta t\rangle$, which is used to obtain the mean velocity $U = Q/\rho Wh$, used in both $Wi = \lambda U/h$ and $Re = \rho Uh/\eta$.

Two-dimensional Particle image velocimetry (PIV) is used in $x-z$ plane to measure the velocity profile and fluctuations. The fluid is seeded with 3.2 $\mu$m particles of $\sim$1\% w/w concentration (Thermo Scientific) and illuminated  by a laser sheet of approximately $50 \mu$m thickness. Images are taken with a high speed, high spatial resolution camera (Mini WX100 FASTCAM, Photron) through a 4X objective. For velocity fluctuation measurements we take a small window of 256$\times$96 px$^2$ as a single point at the channel center, while for velocity profile measurements we enlarge the window up to 1280$\times$1280 px$^2$. The window size for PIV is $32\times 32$ px$^2$ with 50\% overlap with 200\% search window size. The OpenPIV software [1] is used to analyze $u(x,z,t)$ and $w(x,z,t)$ in the $x$-$z$ plane to record data for $\sim\mathcal{O}(15)$ minutes, or $\sim\mathcal{O}(50\lambda)$, for each $Wi$ to obtain sufficient statistics.

The working fluid is a dilute solution of the high molecular weight polyacrylamide (PAAm) with $M_w=18$ MDa. The Newtonian solvent consists of $64\%$ sucrose and $1\%$ NaCl dissolved in water. By adding a PAAm concentration of $c=80$ ppm ($c/c*\approx0.4$ with the overlap PAAm concentration $c*\approx200$ ppm [2]) to the solvent, we obtain viscoelastic working fluid. The properties of the solution include a density of $\rho=1320$ kg/m$^3$, and solvent and solution viscosities of $\eta_s=0.13$ Pa$\cdot$s and $\eta = \eta_s+\eta_p=0.17$ Pa$\cdot$s, respectively,  $\eta_s/(\eta_s+\eta_p)=0.765$, where $\eta_p$ is the polymer contribution to the solution viscosity, and the longest polymer relaxation time of $\lambda=13$ s, obtained by the stress relaxation method [2].

\pagebreak
 
\subsection*{Supplementary figures}

\setcounter{figure}{0} % to restart counting figure numbers
\renewcommand{\figurename}{Fig. S} % to rename the names of figure in Extended Data

\begin{figure}[h]
\centering
\includegraphics{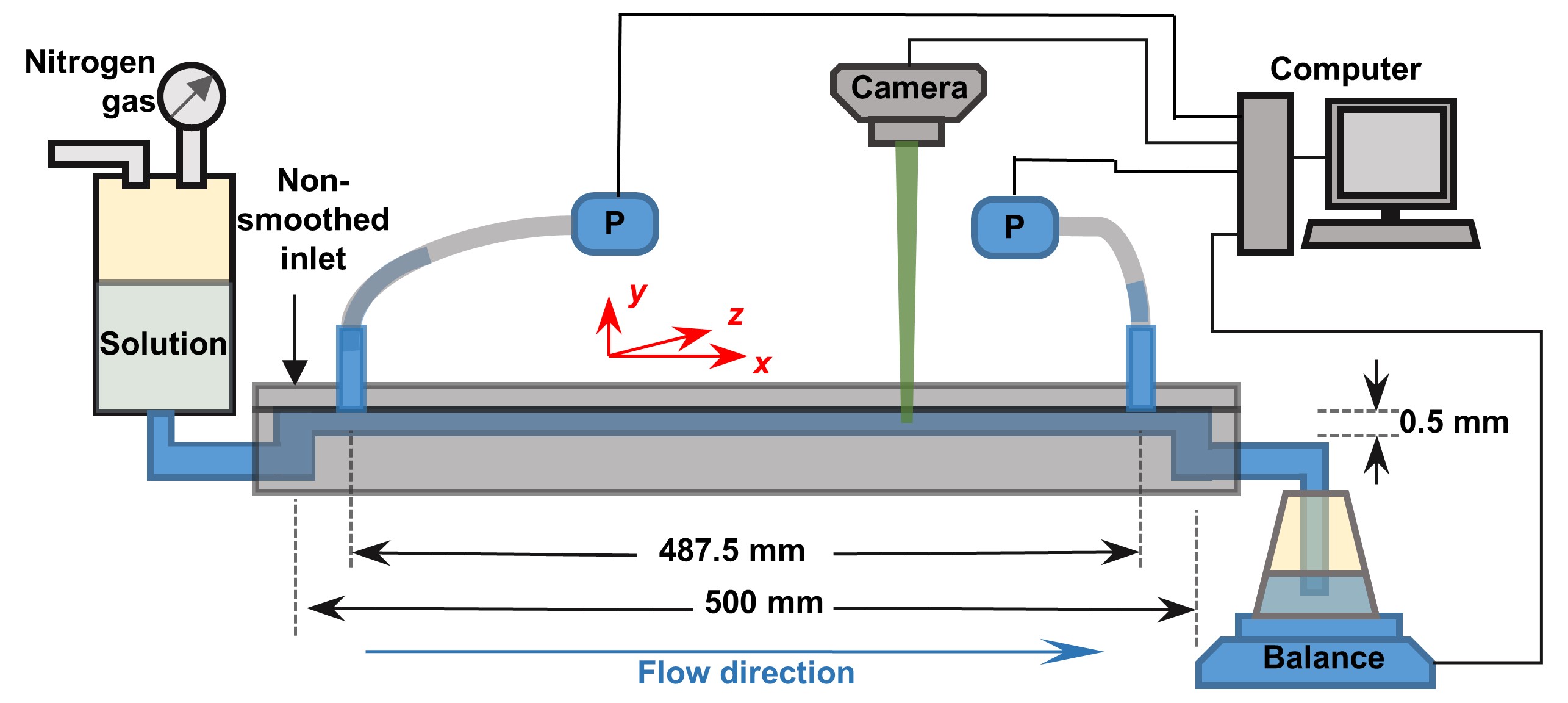}
\caption{\label{fig1:schematics}
Schematic of the experimental set-up. The polymer solution is stored in a large stainless steel vessel and is driven by compressed nitrogen gas controlled by pressure regulators. The solution flows through a straight channel with a unsmoothed inlet and two small holes in the top wall. The solution exiting the outlet is collected in a beaker on the balance. The two holes provide the pressure measurement, and the balance measures the flow discharge rate. Seeded tracers are illuminated by a thin laser sheet along $x-z$ plane at the center of the channel and a high speed camera is used for image acquisition. }
\end{figure}

\begin{figure}
\includegraphics{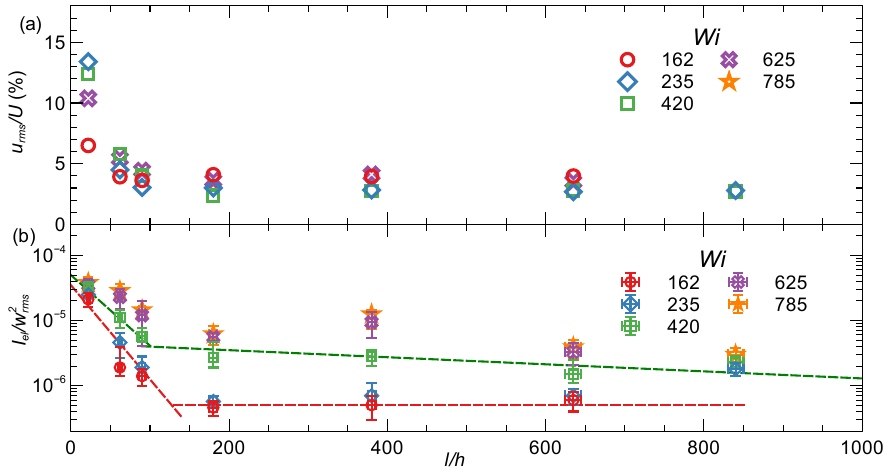}
  \caption{\label{fig2:urms}(a) Normalized streamwise velocity fluctuations, $u_{rms}/U$, and (b) elastic wave energy, $I_{el}$ normalized by the square of the spanwise velocity fluctuations, $I_{el}/w_{rms}^{2}$, versus different downstream locations from the inlet, $l/h$, at five different $Wi$ in the transition flow regime. The red dashed lines in (b) show the decay for the lower sub-range, while the green lines show the decay for the upper sub-range.}
\end{figure}

\begin{figure}
\includegraphics{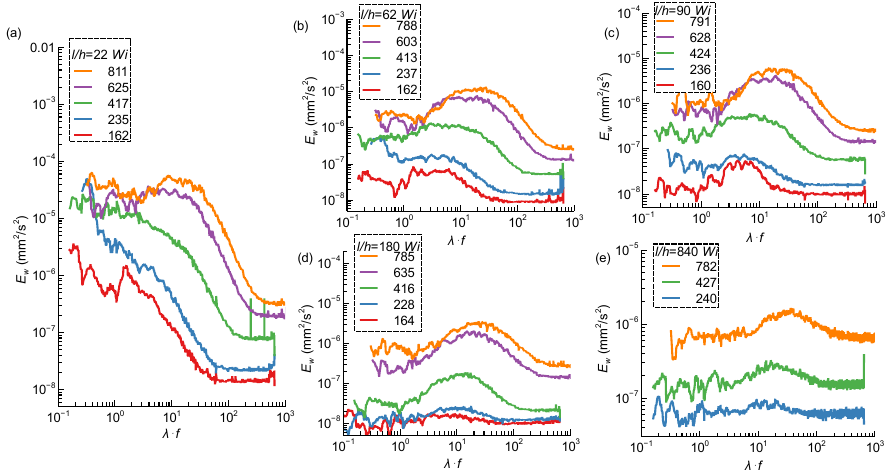}
  \caption{\label{fig4:Ew} Spanwise velocity spectrum, $E_{w}$, versus normalized frequency, $\lambda\cdot{f}$, at five downstream locations, $l/h$=22, 62, 90, 180 and 840, and several $Wi$ values in the transition flow regime. }
\end{figure}

\begin{figure}
\includegraphics{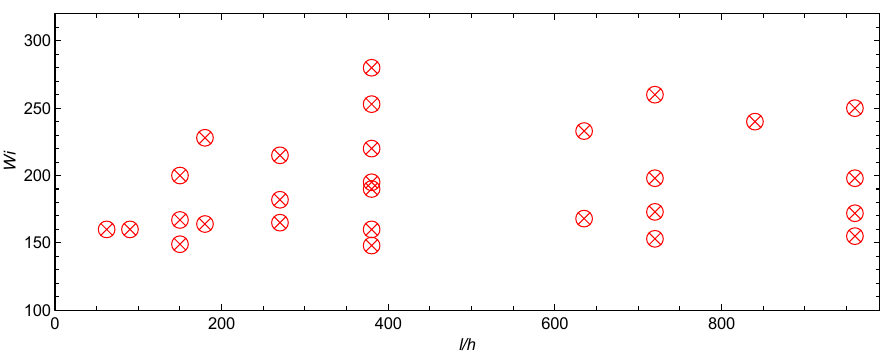}
  \caption{\label{fig3:subrange} Stochastic resonance peaks found in the lower sub-range are shown in $l/h$ and $Wi$ coordinates. }
\end{figure}

\begin{figure}
\includegraphics{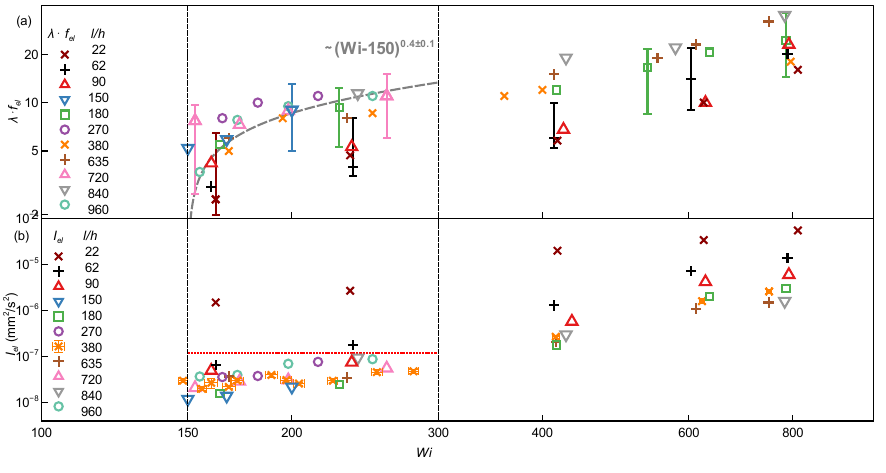}
  \caption{\label{fig:fel}(a) Normalized elastic wave frequency, $\lambda\cdot{f_{el}}$, and (b) elastic wave energy, $I_{el}$, versus $Wi$ at several downstream locations, $l/h=$22, 62, 90, 180, 270, 380, 635, 720, 840, and 960. $\lambda\cdot{f_{el}}$ is plotted with selected error bars. The dashed line is the power-law fit  $\lambda\cdot{f_{el}}\sim(Wi-150)^{\beta}$ with $\beta=0.4\pm 0.1$. The vertical black dashed lines indicate  $Wi_c$=150, and $Wi$=300, the approximate upper limit of the lower subrange of the transition flow regime, where the SR energy, $I_p$, exceeds the elastic wave energy, $I_{el}$, by up to two orders of magnitude and drives chaotic flow. The horizontal red dotted line in (b) indicates the limit above which the elastic wave energy is sufficient to take over.}
\end{figure}
\newpage

\subsection*{Supplementary table}
Table S1: Table comparing the streamwise spike intensity, $I_p$, and the spanwise elastic wave intensity, $I_{el}$. The rows in light green are for the lower sub-range, while the rows in light yellow are for the upper sub-range in the transition regime. $I_{el}$ is also calculated on the smoothed spanwise spectrum because the elastic wave peaks are broad and noisy. Plots of these data are also shown in Fig. 4(b), Fig. S2(b) and Fig. S5(b).
\begin{longtable}{lllll}
\hline \hline
\rowcolor[HTML]{EDEDED} 
$l/h$ & $Wi$  & $I_p$ ($mm^2/s^2$) & \multicolumn{2}{l}{\cellcolor[HTML]{EDEDED}$I_{el}$ ($mm^2/s^2$)} \\
\rowcolor[HTML]{EDEDED} 
    &     & raw         & raw                        & averaged                    \\ \hline
\rowcolor[HTML]{FFF2CC} 
22  & 162 &             & 5.6e-6                     & 1.5e-6                      \\
\rowcolor[HTML]{FFF2CC} 
22  & 235 &             & 5.7e-6                     & 2.7e-6                      \\
\rowcolor[HTML]{FFF2CC} 
22  & 417 &             & 3.7e-5                     & 2e-5                        \\
\rowcolor[HTML]{FFF2CC} 
22  & 625 &             & 1.8e-4                     & 3.4e-5                      \\
\rowcolor[HTML]{FFF2CC} 
22  & 811 &             & 4e-4                       & 5.5e-5                      \\
\rowcolor[HTML]{E2EFD9} 
%62  & 162 & 6e-5        & 3.3e-7                     & 6.5e-8                      \\
62  & 162 & 5e-5        & 3.3e-7                     & 6.5e-8                      \\
\rowcolor[HTML]{FFF2CC} 
62  & 237 &             & 8.6e-7                     & 1.8e-7                      \\
\rowcolor[HTML]{FFF2CC} 
62  & 413 &             & 6.1e-6                     & 1.3e-6                      \\
\rowcolor[HTML]{FFF2CC} 
62  & 603 &             & 3.5e-5                     & 7.2e-6                      \\
\rowcolor[HTML]{FFF2CC} 
62  & 788 &             & 8e-5                       & 1.4e-5                      \\
\rowcolor[HTML]{E2EFD9} 
90  & 160 & 1.4e-4      & 3.6e-7                     & 5e-8                        \\
%\rowcolor[HTML]{E2EFD9} 
%90  & 236 & 7e-5        & 4.2e-7                     & 7.5e-8                      \\
\rowcolor[HTML]{FFF2CC}
90  & 236 &             & 4.2e-7                     & 7.5e-8                      \\
\rowcolor[HTML]{FFF2CC} 
90  & 434 &             & 3.3e-6                     & 5.8e-7                      \\
\rowcolor[HTML]{FFF2CC} 
90  & 628 &             & 2.1e-5                     & 4.2e-6                      \\
\rowcolor[HTML]{FFF2CC} 
90  & 791 &             & 3.5e-5                     & 6e-6                        \\
\rowcolor[HTML]{E2EFD9} 
150 & 150 & 1.4e-6      & 5.6e-8                     & 1.2e-8                      \\
\rowcolor[HTML]{E2EFD9} 
150 & 167 & 8.6e-6      & 9e-8                       & 1.4e-8                      \\
\rowcolor[HTML]{E2EFD9} 
150 & 200 & 7.8e-5      & 1.5e-7                     & 2.2e-8                      \\
\rowcolor[HTML]{E2EFD9} 
180 & 164 & 2e-5        & 1e-7                       & 1.6e-8                      \\
\rowcolor[HTML]{E2EFD9} 
180 & 228 & 2.7e-5      & 1.5e-7                     & 2.5e-8                      \\
\rowcolor[HTML]{FFF2CC} 
180 & 416 &             & 1.1e-6                     & 1.8e-7                      \\
\rowcolor[HTML]{FFF2CC} 
180 & 635 &             & 1.1e-5                     & 2e-6                        \\
\rowcolor[HTML]{FFF2CC} 
180 & 785 &             & 2.1e-5                     & 3e-6                        \\
\rowcolor[HTML]{E2EFD9} 
270 & 165 & 1e-4        & 2.3e-7                     & 3.6e-8                      \\
\rowcolor[HTML]{E2EFD9} 
270 & 182 & 1e-5        & 2.8e-7                     & 3.8e-8                      \\
\rowcolor[HTML]{E2EFD9} 
270 & 215 & 2e-5        & 5e-7                       & 7.7e-8                      \\
\rowcolor[HTML]{E2EFD9} 
380 & 148 & 7.6e-7      & 1.6e-7                     & 3e-8                        \\
\rowcolor[HTML]{E2EFD9} 
380 & 156 & 1e-6        & 1.1e-7                     & 2e-8                        \\
\rowcolor[HTML]{E2EFD9} 
380 & 160 & 3.4e-6      & 1.2e-7                     & 2.7e-8                      \\
\rowcolor[HTML]{E2EFD9} 
380 & 168 & 7.2e-6      & 8.8e-8                     & 2.2e-8                      \\
\rowcolor[HTML]{E2EFD9} 
380 & 172 & 1.5e-5      & 1.6e-7                     & 3e-8                        \\
\rowcolor[HTML]{E2EFD9} 
380 & 189 & 2.4e-5      & 2.5e-7                     & 4e-8                        \\
\rowcolor[HTML]{E2EFD9} 
380 & 197 & 1.9e-5      & 1.8e-7                     & 3.1e-8                      \\
\rowcolor[HTML]{E2EFD9} 
380 & 204 & 4.6e-5      & 1.4e-7                     & 2.6e-8                      \\
\rowcolor[HTML]{E2EFD9} 
380 & 224 & 6.2e-5      & 1.6e-7                     & 3e-8                        \\
\rowcolor[HTML]{E2EFD9} 
380 & 253 & 1.4e-5      & 2.7e-7                     & 4.6e-8                      \\
\rowcolor[HTML]{E2EFD9} 
380 & 280 & 2e-5        & 3e-7                       & 4.8e-8                      \\
\rowcolor[HTML]{FFF2CC} 
380 & 415 &             & 1.6e-6                     & 2.7e-7                      \\
\rowcolor[HTML]{FFF2CC} 
380 & 622 &             & 8.8e-6                     & 1.6e-6                      \\
\rowcolor[HTML]{FFF2CC} 
380 & 749 &             & 1.5e-5                     & 2.6e-6                      \\
\rowcolor[HTML]{E2EFD9} 
635 & 168 & 9.5e-6      & 2.9e-7                     & 3.7e-8                      \\
\rowcolor[HTML]{E2EFD9} 
635 & 233 & 7.2e-6      & 2.7e-7                     & 3.5e-8                      \\
\rowcolor[HTML]{FFF2CC} 
635 & 415 &             & 1.2e-6                     & 2.1e-7                      \\
\rowcolor[HTML]{FFF2CC} 
635 & 612 &             & 9e-6                       & 1.1e-6                      \\
\rowcolor[HTML]{FFF2CC} 
635 & 748 &             & 1.2e-5                     & 1.5e-6                      \\
\rowcolor[HTML]{E2EFD9} 
720 & 153 & 4.5e-6      & 1.2e-7                     & 2.1e-8                      \\
\rowcolor[HTML]{E2EFD9} 
720 & 173 & 1.2e-6      & 1.6e-7                     & 2.9e-8                      \\
\rowcolor[HTML]{E2EFD9} 
720 & 198 & 7e-6        & 1.8e-7                     & 3.1e-8                      \\
\rowcolor[HTML]{E2EFD9} 
720 & 260 & 3e-5        & 3.5e-7                     & 5.6e-8                      \\
\rowcolor[HTML]{E2EFD9} 
840 & 240 & 7.3e-6      & 9.6e-8                     & 9.4e-8                      \\
\rowcolor[HTML]{FFF2CC} 
840 & 427 &             & 1.9e-6                     & 3e-7                        \\
\rowcolor[HTML]{FFF2CC} 
840 & 782 &             & 1e-5                       & 1.6e-6                      \\
\rowcolor[HTML]{E2EFD9} 
960 & 155 & 2e-5        & 2e-7                       & 3.7e-8                      \\
\rowcolor[HTML]{E2EFD9} 
960 & 172 & 3.7e-5      & 2.7e-7                     & 4e-8                        \\
\rowcolor[HTML]{E2EFD9} 
960 & 198 & 4e-6        & 3.5e-7                     & 7e-8                        \\
\rowcolor[HTML]{E2EFD9} 
960 & 250 & 2e-5        & 4.5e-7                     & 8.8e-8                     \\ \hline \hline
\end{longtable}

\subsection*{References}
[1] A. Liberzon, T. Käufer, A. Bauer, P. Vennemann and E. Zimmer. OpenPIV/openpiv-python: OpenPIV-Python v0.23.3. 10.5281/zenodo.4320056 (Zenodo, 2020)

[2] Y. Liu, Y.  Jun and V. Steinberg. Concentration dependence of the longest relaxation times of dilute and semi-dilute polymer solutions. J. Rheol. \textbf{53} 1069 (2009)

%\nocite{*}
%\bibliography{bibliography.bib}